\documentstyle[12pt]{article}

\renewcommand{\thefootnote}{\fnsymbol{footnote}}
\newcommand{\newsection}{
\setcounter{equation}{0}
\section}

\def\appendix#1{
  \addtocounter{section}{1}
  \setcounter{equation}{0}
  \renewcommand{\thesection}{\Alph{section}}
  \section*{Appendix \thesection\protect\indent \parbox[t]{11.715cm} {#1}}
  \addcontentsline{toc}{section}{Appendix \thesection\ \ \ #1}
  }

\def\E{\sim}

\newcommand{\tr}[1]{\:{\rm Tr}\,#1}

\def\e{{\,\rm e}\,}
\def\eol{\hspace*{\fill}\linebreak}
\def\eop{\vspace*{\fill}\pagebreak}
\newcommand{\rf}[1]{(\ref{#1})}
\newcommand{\eq}[1]{Eq.~(\ref{#1})}
\def\be{\begin{equation}}
\def\ee{\end{equation}}
\def\beq{\begin{equation}}
\def\eeq{\end{equation}}
\def\bea{\begin{eqnarray}}
\def\eea{\end{eqnarray}}
\def\LB{\left (}
\def\RB{\right )}
\def\A{{\cal A}}
\def\cl{{\rm cl}}
\def\s{{\mbox{\boldmath $\sigma$}}}
\def\m{{\bf m}}

\newcommand{\non}{\nonumber \\*}

\newcommand{\ra}{\rightarrow}
\def\bl{\Bigl(}

\def\br{\Bigr)}

\begin{document}

\begin{titlepage}
\begin{flushright}
ITEP--TH--07/97\\
NBI--HE--97--01\\
hep-th/9701151\\
January, 1997
\end{flushright}
\vspace{1.5cm}

\begin{center}
{\LARGE Properties of D-Branes in Matrix\\[.4cm]
 Model of IIB Superstring}\\
\vspace{1.9cm}
{\large I. Chepelev, Y. Makeenko%
\footnote{Also at {\it The Niels Bohr Institute,
 Blegdamsvej 17, 2100 Copenhagen \O, Denmark}; \eol
\mbox{} \ \ \ \ E-mail: makeenko@nbi.dk}
and K. Zarembo}\\
\vspace{24pt}
{\it Institute of Theoretical and Experimental Physics,}
\\ {\it B. Cheremushkinskaya 25, 117259 Moscow, Russia} \\ \vskip .5 cm
E-mail: {\tt makeenko,zarembo@vxitep.itep.ru}
\end{center}
\vskip 2 cm
\begin{abstract}
We discuss properties of D-brane configurations in the
matrix model of type IIB superstring
recently proposed
by Ishibashi, Kawai, Kitazawa and Tsuchiya.
We calculate central charges in supersymmetry algebra
at infinite N and associate them with one- and five-branes
present in IIB superstring theory.
We consider classical solutions associated with
static three- and five-branes and calculate
their interactions at one loop in the matrix model.
We discuss some aspects of the
matrix-model formulation of IIB superstring.
\end{abstract}

\end{titlepage}
\setcounter{page}{2}
\renewcommand{\thefootnote}{\arabic{footnote}}
\setcounter{footnote}{0}

\newsection{Introduction}

It has been recently proposed by Banks, Fischler, Shenker and
Susskind~\cite{BFSS96} that nonperturbative dynamics of M~theory
is described by a supersymmetric $N\times N$  matrix
quantum mechanics in the limit of large $N$.
This Matrix theory naturally includes Witten's description~\cite{Wit95}
of bound states of D(irichlet)-branes by matrices and is shown
\cite{BFSS96,grt,AB96,LM96,bss,Lif96} to correctly
reproduce properties of Dp-branes with even $p$ ($p=0,2,4,\ldots$)
incorporated by type IIA superstring theory.

Another matrix model which is an analogue of the BFSS
matrix model~\cite{BFSS96} for type IIB superstring
has been proposed by Ishibashi, Kawai, Kitazawa and
Tsuchiya~\cite{ikkt}.
This model is defined by the vacuum amplitude
\beq
Z= \sum_{n=1}^{\infty} \int dA \,d\psi \, \e^{iS}
\label{main}
\eeq
with the action being
\beq
S=\frac{1}{g_s(\alpha')^2}\left( \frac{1}{4}\tr[A_{\mu},A_{\nu}]^2
            +\frac{1}{2}\tr (\bar{\psi}
           \Gamma^{\mu}[A_{\mu},\psi])\right)+\beta n .
\label{action}
\eeq
Here $A_{\mu}^{ij}$ and $\psi_\alpha^{ij}$ are
$n \times n$ Hermitian bosonic and fermionic matrices, respectively.
The vector index $\mu$ runs from $0$ to $9$ and the spinor
index $\alpha$  runs
from $1$ to $32$. The fermion $\psi$ is a Majorana--Weyl
 spinor which satisfies the condition $\Gamma_{11}\psi=\psi$.
The summation over $\mu$ is understood with
ten-dimensional Minkowski metric. We prefer to
 work with the vacuum amplitude in Minkowski space rather
than with Euclidean partition function to
avoid problems with  Majorana--Weyl spinors in Euclidean space.

The action~\rf{action} is invariant under
the ${\cal N}=2$ supersymmetry transformations
\bea
\delta^{(1)}\psi^{ij}_\alpha &=& \frac{i}{2}
           [A_{\mu},A_{\nu}]^{ij}(\Gamma^{\mu\nu}\epsilon)_\alpha ,\non
\delta^{(1)} A_{\mu}^{ij} &=& i\bar{\epsilon}\Gamma_{\mu}\psi^{ij} ,
\label{sym1}
\eea
and
\bea
\delta^{(2)}\psi^{ij}_\alpha &=& \xi_\alpha \delta^{ij} ,\non
\delta^{(2)} A_{\mu}^{ij} &=& 0.
\label{sym2}
\eea
The formulas look like as if ten-dimensional super Yang-Mills
theory is reduced to a point%
\footnote{Another matrix model on a point was advocated
in~\cite{Per96}.}.

The type IIB superstring theory consistently incorporates~\cite{pol}
Dp-branes with odd $p$ ($p=-1,1,3,5,\ldots$). In order for the matrix
model to describe nonperturbative dynamics of type IIB string, it
should correctly reproduce the central charges in the supersymmetry
algebra. These central charges have nontrivial tensor structure and
are associated with D-branes of various dimensions.

The action \rf{action} is, up to a constant term, the low energy effective
 action of \mbox{D-instanton} (associated with $p=-1$) \sloppy
 of charge $n$~\cite{Wit95}.
Higher dimensional branes are expected to
 show up in the matrix model as solutions
of the classical equations
\be
\left[ A^{\mu}\left[ A_\mu,A_\nu \right]\right]=0\,,~~~~~
\left[ A_\mu\,, (\Gamma^\mu\psi)_\alpha \right] =0\,,
\label{ce}
\ee
which are to be solved for $n\times n$ matrices ${A}_\mu$
at infinite $n$.
A general solution has a block-diagonal form and is composed from  
non-diagonal $n_i\times n_i$ matrices with various $n_i$.
A simplest solution corresponds to a diagonal matrix
\be
{A}_\mu^\cl= \hbox{diag}\left(p_\mu^{(1)}, \ldots,
p_\mu^{(n)} \right),~~~~~\psi_\alpha=0\,.
\label{D-instanton}
\ee
In analogy with Ref.~\cite{BFSS96}, each of $p_\mu$'s is to
be identified with the coordinates of
D-instanton
which generates space-time coordinates.

As is discussed in~\cite{ikkt},
D-strings (associated with $p=1$) are also described by 
the matrix model \rf{main}
using the idea of Ref.~\cite{BFSS96} to identify D-branes
with operator-like solutions of \eq{ce}.
A static D-string extending along the $\mu=1$ axis is
represented by
\be
A_\mu^\cl=\left(B_0,B_1,0,\ldots,0\right),~~~~~\psi_\alpha^{\cl}=0\,,
\label{D-string}
\ee
where the operators (infinite $n\times n$ matrices)
$B_0$ and $B_1$ obey canonical commutation relation on a torus.
The torus is associated with large compactification radii, $T$
and $L$, along the $\mu=0,1$ directions so that the ratio
$TL/n=\alpha^\prime$ is kept fixed as $n\ra\infty$.
As is shown in~\cite{ikkt},
the interaction between two classical solutions~\rf{D-string},
calculated at the one loop level in the matrix model~\rf{main},
agrees with that of D-strings in IIB supergravity. This confirms
the identification of the classical solution~\rf{D-string} with
D-string. The emergence of the Born--Infeld action
has been also discussed~\cite{Li96}.

In the present paper we consider how
three- and five-branes are described by the matrix model~\rf{main}.
In Sect.~2 we calculate central charges in supersymmetry algebra
at infinite $n$ and associate them with one- and five-branes
present in IIB superstring theory.
In Sect.~3 we consider classical solutions associated with
three- and five-branes and calculate
their interactions at one loop in the matrix model.
In Sect.~4 we give a general prescription for taking
the large $n$ limit appropriate for the description of Dp-branes in
the IKKT matrix model.
Finally we discuss in Sect.~5 some aspects of the
matrix-model formulation of IIB superstring.

\newsection{Central charges in supersymmetry algebra}

The supersymmetry transformations~\rf{sym1} and \rf{sym2}, under which
the action \rf{action} is invariant, are
generated by two supercharges
 \begin{equation}\label{Q1}
 Q^{(1)}_{\alpha }=-\frac{i}{2}\,[A_{\mu },A_{\nu }]^{ij}
 \Gamma ^{\mu \nu }_{\alpha \beta }\,
 \frac{\partial }{\partial \bar{\psi }_{\beta}^{ij}}
 +i\Gamma _{\mu \alpha \beta }\psi _{\beta}^{ij}\,\frac{\partial }{\partial
 A_{\mu}^{ij}}
 \end{equation}
 and
 \begin{equation}\label{Q2}
 Q^{(2)}_{\alpha }=\frac{\partial }{\partial \bar{\psi
 }_{\alpha}^{ii}}.
 \end{equation}
 These operators form the ${\cal N}=2$ supersymmetry algebra \cite{ikkt},
which is not central extended at finite $n$.

The situation changes at $n=\infty$.
 As was shown in \cite{bss}, the supersymmetry algebras in matrix
 models can acquire central charges in the infinite $n$ limit. This happens
 because the quantities proportional to the traces of commutators, which
 vanish for finite matrices and are usually dropped in the calculation
 of the anticommutation relations in supersymmetry algebra, can be not
 equal to zero for operators in the Hilbert space.
 If matrix commutators are replaced in the large $n$ limit by
 Poisson brackets and the traces are substituted
by the integrals over parameter space, the trace of the
 commutator takes the form of an integral of the full derivative
what is typical
 for central charges. Such terms should be retained and lead to the
 central extension of the supersymmetry algebra. In the
BFSS matrix model they
 were calculated in~\cite{bss}. We shall perform the analogous
 calculation for the IKKT matrix model.

 We introduce the operators
 \begin{equation}\label{defP}
 P^{\mu }_{ij}=\frac{\partial }{\partial A_{\mu}^{ji}},
 \end{equation}
 \begin{equation}\label{defch}
 \bar{\chi} _{\alpha ij}=\frac{\partial }{\partial \psi _{\alpha}
 ^{ji}},~~~~
 \chi _{\alpha ij}=-\,\frac{\partial }{\partial \bar{\psi }_{\alpha}^{ji}}.
 \end{equation}
 Note that $\psi $ and $\chi $ have opposite chirality. We shall denote
 (anti)commutators of differential operators by $[\:,]_{\pm}$,
 while for matrices we shall use the symbols $\{\,,\}$ and
 $[\:,]$. We follow the convention that matrix (anti)commutators
 do not change an operator ordering. For example,
\be
 [A,B]_{ij}\equiv A_{ik}B_{kj}-A_{kj}B_{ik}.
\ee
 In this notations, the generator of the infinitesimal
gauge transformation,
\bea
\delta_{{\rm gauge}} A_{\mu} &=& i \left[A_\mu,\Omega\right], \non
\delta_{{\rm gauge}} \psi_{\alpha} &=& i \left[\psi_\alpha,\Omega\right]
\label{gauge}
\eea
reads
 \begin{equation}\label{fi}
 \Phi_{ij} =[A_{\mu },P^{\mu }]_{ij}-[\bar{\psi },\chi ]_{ij}\,.
 \end{equation}

 We define the operators
 \begin{equation}\label{q1}
 q^{(1)}_{\alpha ij}=\frac{i}{4}\,\{[A_{\mu },A_{\nu }],
 (\Gamma ^{\mu \nu }\chi )_{\alpha }\}_{ij}
 +\frac{i}{2}\,\{(\Gamma _{\mu }\psi )_{\alpha },P^{\mu }\}_{ij}
 \end{equation}
 and
 \begin{equation}\label{q2}
 q^{(2)}_{\alpha ij}=-\chi _{\alpha ij},
 \end{equation}
 which are the counterparts of supercharge densities in the BFSS matrix
model, since
 \begin{equation}\label{Q=trq}
 Q^{(1),(2)}_{\alpha }=\tr q^{(1),(2)}_{\alpha }.
 \end{equation}

 To find the central charges in supersymmetry algebra, we first calculate
 the anticommutators of the densities and the
 supercharges. In this rather lengthy calculation we use the
 following Fierz identity for
 ten-dimensional Majorana--Weyl spinors:
 \begin{eqnarray}\label{fierz}
 (\Gamma _{\mu }\psi )_{(\alpha }\otimes(\Gamma ^{\mu \nu }\chi )_{\beta
 )}&=&2(\Gamma ^{\nu }\Gamma ^0)_{\alpha \beta }\bar{\psi }\otimes\chi
 -\,\frac{7}{8}\,(\Gamma ^{\mu }\Gamma ^0)_{\alpha \beta }\bar{\psi
 }\Gamma ^{\nu }\Gamma _{\mu }\otimes\chi
 \non &&
 +\,\frac{1}{8\cdot 5!}\,(\Gamma
 ^{\mu \lambda \rho \sigma \tau }\Gamma ^0)_{\alpha \beta }\bar{\psi
 }\Gamma ^{\nu }\Gamma _{\mu \lambda \rho \sigma \tau }\otimes\chi.
 \end{eqnarray}
 Using this formula we finally obtain
 \begin{eqnarray}\label{antic}
 {[q^{(2)}_{\alpha ij},Q_{\beta }^{(2)}]}_+ & = & 0,\non
 {[q^{(1)}_{\alpha ij},Q_{\beta }^{(2)}]}_+ & = & -i(\Gamma _{\mu }\Gamma
 ^0)_{\alpha \beta }P^{\mu }_{ij},              \non
 {[q^{(1)}_{(\alpha ij},Q_{\beta) }^{(1)}]}_{+} & = &
 2(\Gamma ^{\mu }\Gamma ^0)_{\alpha \beta
 }\,z_{\mu ij}+2(\Gamma ^{\mu \nu \lambda \rho \sigma }\Gamma ^0)_{\alpha
 \beta }\,z_{\mu \nu \lambda \rho \sigma ij}\non &&-2(\Gamma ^{\mu }\Gamma
 ^0)_{\alpha \beta }\{A_{\mu },\Phi \}_{ij} +\,\frac{7}{8}\,(\Gamma ^{\mu
 }\Gamma ^0)_{\alpha \beta }\{[\bar{\psi },A_{\nu }]\Gamma ^{\nu }\Gamma
 _{\mu },\chi \}_{ij}
 \non &&
 -\,\frac{1}{8\cdot 5!}(\Gamma ^{\mu \lambda \rho
 \sigma\tau  }\Gamma ^0)_{\alpha \beta }\{[\bar{\psi },A_{\nu }]\Gamma
 ^{\nu }\Gamma _{\mu \lambda \rho \sigma \tau },\chi \}_{ij},
 \end{eqnarray}
 where
 \begin{eqnarray}\label{centc}
 z_{\mu }&=&[A_{\nu },\{A_{\mu },P^{\nu }\}]-[\bar{\psi },A_{\mu }\chi ]
 -[A_{\mu }\bar{\psi },\chi ]+\,\frac{7}{32}\,[\bar{\psi }A_{\nu }\Gamma
 ^{\nu }\Gamma _{\mu },\chi ],
 \non
 z_{\mu \lambda \rho \sigma \tau }&=&-\,\frac{1}{32\cdot 5!}\,[\bar{\psi
 }A_{\nu }\Gamma ^{\nu }\Gamma _{\mu \lambda \rho \sigma \tau },\chi ].
 \end{eqnarray}

 Taking the trace of \eq{antic}, we find that, up to the gauge
 transformations and equations of motion for $\bar{\psi }$, the
 supercharges obey the anticommutation relations
 \begin{eqnarray}\label{anticc}
 {[Q^{(2)}_{\alpha },Q_{\beta }^{(2)}]}_+ & = & 0,\nonumber \\
 {[Q^{(1)}_{\alpha},Q_{\beta }^{(2)}]}_+ & = & -i(\Gamma _{\mu }\Gamma
 ^0)_{\alpha \beta }\tr P^{\mu },              \non
 {[Q^{(1)}_{\alpha},Q_{\beta}^{(1)}]}_{+} & = &
 (\Gamma ^{\mu }\Gamma ^0)_{\alpha \beta
 }Z_{\mu}+(\Gamma ^{\mu \nu \lambda \rho \sigma }\Gamma ^0)_{\alpha
 \beta }Z_{\mu \nu \lambda \rho \sigma}.
 \end{eqnarray}
 The central charges, $Z_{\mu }=\tr z_{\mu }$ and $Z_{\mu \nu \lambda
 \rho \sigma}=\tr z_{\mu
 \nu \lambda \rho \sigma}$, being equal to the traces of the commutators,
 vanish for finite $n$. But at $n=\infty$ they are not necessarily turn
 to zero and we associate them with one- and five-branes present
in type IIB superstring theory.

It is worth mentioning that all the charges are
 operator-valued and their interpretation is not as clear
as for those of Ref.~\cite{bss} in the BFSS matrix model,
where the value of the charges is given by substituting the
classical solution.  Also there is no three-brane charge in
the supersymmetry algebra.  Similarly, the five-brane charge has
purely fermionic nature.  This circumstance may cause difficulties in
the description of three- and five-branes as certain classical field
configurations of the matrix model.  Nevertheless, in the next
section we shall study some classical solutions of the matrix model,
which can be seemingly interpreted as D-branes of different
dimensions.

\newsection{Brane--brane interaction}

 It was argued in \cite{bss,ikkt} that for BPS states the field strength
 \begin{equation}\label{fmn}
 f_{\mu \nu }=i[A_{\mu },A_{\nu }],
 \end{equation}
 should be proportional
to the unit matrix. The classical equations~\rf{ce}
are in this case automatically satisfied.
Since D-branes are BPS-states~\cite{pol},
classical solutions of the matrix model which
correspond to D-branes should have this property.

Motivated by the four-brane solution found in~\cite{bss}
for the BFSS matrix model, we
associate with a static Dp-brane the following classical
solution of the model~\rf{action}:
\be
A_\mu^\cl = \left(B_0,B_1,B_2,\ldots,B_p,0,\ldots,0   \right),
~~~~~\psi_\alpha^\cl=0\,,
\label{Dp}
\ee
where $B_0,\ldots,B_p$ are operators (infinite matrices)
with the commutator
\be                        \label{bb}
[B_a,B_b]=-ig_{ab}{\bf 1}\,,
\ee
where $a,b=0,\ldots,p$.
Since $p$ is odd, these $B_a$'s can be written
as linear combinations of $(p+1)/2$ pairs of canonical variables
$p_k,q_k$ ($k=1,\ldots,(p+1)/2$) satisfying
$[q_k,p_l]=i\delta_{kl}$. The solution~\rf{Dp} is an obvious
extension of~\rf{D-string}. The property~\rf{bb} guarantees that the
effective action in the background~\rf{Dp} does not acquire quantum
corrections.

The configuration containing
a pair of Dp-branes can then be constructed embedding the
 classical solutions~\rf{Dp}
in $A_{\mu }$ diagonally. We shall study in this section most
 general background configurations of this type, which are very similar
 to the one considered in the context of the BFSS matrix model
 in \cite{Lif96} and
 generalize the configurations with two static
D-strings of Ref.~\cite{ikkt}.

 The natural choice of the classical solution which can be interpreted as
 two parallel (antiparallel)
Dp-branes at the distance $b$ from each other is
 \begin{eqnarray}\label{backgr}
 A_{a}^\cl&=&\left(
 \begin{array}{cc}
 B_a & 0 \\
 0 & B'_a \\
 \end{array}
 \right),~~~~a=0,\ldots,p\non \label{backgr1}
 A_{p+1}^\cl&=&\left(
 \begin{array}{cc}
 b/2 & 0 \\
 0 & -b/2 \\
 \end{array}
 \right),\non    \label{backgr2}
 A_{i }^\cl&=&0,~~~~i =p+2,\ldots,9,
 \end{eqnarray}
 where
 \begin{equation}\label{qp}
 [B_a,B_b]=-ig_{ab}{\bf 1},~~~~
 [B'_a,B'_b]=-ig'_{ab}{\bf 1}.
 \end{equation}

 For this configuration
 \begin{equation}\label{fmndp}
 f_{ab}=\left(
 \begin{array}{cc}
 g_{ab} & 0 \\
 0 & g'_{ab} \\
 \end{array}
 \right)=d_{ab}\otimes{\bf 1}_2+c_{ab}\otimes\sigma ^3,
 \end{equation}
 \begin{eqnarray}\label{defdc}
 d_{ab}&=&\frac{g_{ab}+g'_{ab}}{2},\\
 c_{ab}&=&\frac{g_{ab}-g'_{ab}}{2},
 \end{eqnarray}
 and all other $f_{\mu \nu }$ are equal to zero.
The matrix $c_{ab}$ can
 always be brought to the canonical form by Lorentz transformation, so
 without loss of generality we can assume that
 \begin{equation}\label{stc}
 c_{ab}=\left(
 \begin{array}{ccccc}
 0 & -\omega_1 & & & \\
 \omega_1 &0& & & \\
  & &\ddots & & \\
 &&&0 & -\omega_{\frac{p+1}{2}}  \\
 &&&\omega_{\frac{p+1}{2}} &0 \\
 \end{array}
 \right).
 \end{equation}

If the Dp-branes are parallel, we can set $B_a=B'_a$
by a canonical transformation, so that $c_{ab}=0$.
This corresponds again to the BPS-saturated case.
If the Dp-branes are antiparallel, say, along one axis,
then $c_{ab}\neq0$ and their
interaction is to be calculated.
We use for this purpose the result of Ref.~\cite{ikkt}
for the one-loop effective action around a general background
$A_\mu^\cl$ satisfying \eq{ce}:
\bea\label{seff}
W&=&\frac{1}{2}{\rm Tr}\ln(P^2\delta_{\mu
\nu}-2iF_{\mu\nu})-\frac{1}{4}{\rm Tr}\ln\LB (
P^2+\frac{i}{2}F_{\mu\nu}\Gamma^{\mu\nu})\LB\frac{1+\Gamma_{11}}{2}
\RB\RB
\non &&
-{\rm Tr}\ln(P^2),
\eea
where the adjoint operators $P_{\mu}$ and $F_{\mu\nu}$ are defined
on the space of matrices by
\be
P_{\mu}=\left[A_{\mu }^\cl,\cdot\,\right],~~~~~
 F_{\mu\nu}=\left[f_{\mu \nu }^\cl,\cdot\,\right]=
i\left[\left[A_\mu^\cl,A_\nu^\cl\right],\cdot\,\right] .
\label{PF}
\ee
${\rm Im}\:W$ vanishes for $p=1,3,5,7$ since
we have $P_{\mu}=0$ at least in one direction.

The calculation of~\rf{seff} considerably simplifies
for the background~\rf{backgr2}
when all of the operators $P_\mu$ and $F_{\mu\nu}$
 have the form ${\cal O}_1\otimes
 1+{\cal O}_3\otimes\Sigma^3$ with $\Sigma ^3=[1\otimes\sigma ^3,\cdot\,]$.
 Thus they commute with $\Sigma ^3$ and the eigenfunctions of the
 operators entering~\rf{seff} can be classified according to the
 eigenvalues of $\Sigma ^3$. The terms corresponding to zero eigenvalues
of $\Sigma ^3$
do not contribute to the effective action~\rf{seff}. Two other
 eigenvalues of $\Sigma ^3$ are $\pm 2$ and they give equal contributions.
 The commutation relations
 \begin{equation}\label{oper}
 F_{ab}=i[P_a,P_b]=c_{ab}\Sigma ^3
 \end{equation}
 and the equality
 \begin{equation}\label{pp+1}
 P_{p+1}=\frac{b}{2}\Sigma ^3
 \end{equation}
 show that after analytical continuation to the Euclidean space the
 operator $P^2$ projected on the eigenspace of $\Sigma_3$ can be
 regarded as a Hamiltonian of
 $\left(\frac{p+1}{2}\right)$--dimensional harmonic oscillator with
 frequencies $\omega _i$. Its eigenvalues thus are
 \begin{equation}\label{eigen} E_{\bf
 k}=4\sum_{i=1}^{\frac{p+1}{2}}\omega _i
 \left(k_i+\frac{1}{2}\right)+b^2.
 \end{equation}
 Each of them has $n$--fold degeneracy.

 For
 $c_{ab}$ given by \eq{stc}, the effective action can be brought, as
  was shown in~\cite{ikkt},  to the form
 \begin{equation}\label{seff1}
 W=n\sum_{\bf k}\left[\sum_{i}
 \ln\left(1-\frac{16\omega^2 _i}{E_{\bf k}^2}\right) -\frac{1}{2}\sum_{
 \begin{array}{c}
 \scriptstyle
 s_1,\ldots,s_5=\pm 1\\[-2.5mm]
 \scriptstyle
 s_1\ldots s_5=1
 \end{array}
 }\ln\left(1-\frac{2\sum\limits_{i}\omega _is_i}{E_{\bf k}}\right)\right].
 \end{equation}
 The sums over $k_i$ can be calculated using the formulas
 \begin{equation}\label{log1}
 \ln \frac uv =\int\limits_0^{\infty}\frac{dx}{x}\left(\e^{-vx}
- \e^{-ux} \right)
 \end{equation}
 and
 \begin{equation}\label{log2}
 \sum_{\bf k}\e^{-xE_{\bf k}}=\frac{\e^{-b^2x}}{\prod\limits_i
 2\sinh 2\omega_i x}.
 \end{equation}
 After some algebra we get
 \begin{equation}\label{otvet}
 W=-2n\int\limits_{0}^{\infty }\frac{dx}{x}\,\e^{-b^2x}
 \left[\sum_{i}\bl\cosh 4\omega _ix-1\br-4\bl\prod_{i}\cosh
 2\omega _ix-1\br\right]\prod_{i}\frac{1}{2\sinh 2\omega _ix}.
 \end{equation}
The integral is convergent for $p\leq 5$ and logarithmically
divergent for $p=7$.
 Retaining only the leading term in $1/b^2$, we obtain from
 \eq{otvet}:
 \begin{eqnarray}\label{1/b}
 W&=&-\,\frac{1}{16}\,n\left(\frac{5-p}{2}\right)!\,
 \left[2\sum_{i}\omega _i^4-\left(\sum_{i}\omega _i^2\right)^2\right]
 \prod_{i}\omega _i^{-1}\left(\frac{2}{b}\right)^{7-p}
 +O\left(\frac{1}{b^{9-p}}\right)
 \non
 &=&-\,\frac{1}{64}\,n\left(\frac{5-p}{2}\right)!\,
 \left[4c_{ab}c_{bd}c_{de}c_{ea}-\left(c_{ab}c_{ba}\right)^2\right]
 \left(\det\limits_{ab}c_{ab}\right)^{-1/2}\left(\frac{2}{b}\right)^{7-p}
 \non&&
 +O\left(\frac{1}{b^{9-p}}\right).
\end{eqnarray}

The right-hand side of \eq{1/b} obviously
vanishes for parallel Dp-branes
when \mbox{$c_{ab}=0$} and recovers the result of Ref.~\cite{ikkt} for
$p=1$. For $p=3,5$ it gives a consistent result for
the interaction between two antiparallel Dp-branes which
falls as $1/b^{7-p}$ at large distances, as expected.

It is worth mentioning that for $p=3$ and all $\omega_i$'s equal to each
other the coefficient in \rf{1/b} turns to zero. Moreover, the complete
effective action \rf{otvet} vanishes in this case. However, it does not
mean that antiparallel tree-branes do not interact, because the choice of
 equal $\omega _i$ is not appropriate for studying the interaction
 between branes.
 As is the case of D-strings~\cite{ikkt}, it is natural to put 
\be
 \omega_1=\frac{TL_1}{2\pi n^{\frac{2}{p+1}}}
\ee
 where $L_1$ is  large
  compactification radius in $x_1$ direction and $T$ is the interval of
 time periodicity. Analogously, it is natural to set
 \be
 \omega_i=\frac{L_{2i-2}L_{2i-1}}{2\pi n^{\frac{2}{p+1}}}.
 \ee
 The effective action \rf{otvet} is related to the
 interaction potential when $T\gg L_a$. In this case the coefficient
 before $b^{p-7}$ is always negative, thus the antiparallel branes
 always attract, as they should. The fact
 that
 the effective action vanishes for coinciding radii of
 compactification, although having nothing to do with the interaction
 between D-branes, may have sense, and it would be interesting to find a
 simple explanation of it.

 The correct large distance behaviour of the interaction potential
 confirms the conjecture to identify the solution~\rf{backgr2} with
 Dp-brane configurations.  However, since they are not straightforwardly
 associated, as is already noted, with the central charges calculated in
 the previous section, other checks of this conjecture, in particular the
 derivation of the Born--Infeld action, would be useful.

\newsection{The large $n$ limit for D-branes}

In this section we shall give a prescription for taking the large $n$ limit
appropriate for the description of D-branes in the IKKT matrix
model of IIB string
theory. Analogous prescription in the BFSS matrix model
was given in~\cite{bss}.
Using this prescription, we obtain the correct dependence of physical
quantities, such as the effective action for brane--brane system, on the
fundamental constants.

The action~\rf{action} of the IKKT matrix model is actually
the action for the $p=-1$ D-brane (D-instanton), so the
 higher dimensional D-branes can be viewed as the composites of
 instantons similarly to the BFSS matrix model, in which D-branes are
 composed from D0-branes.
 In \cite{BFSS96} it was shown that
the transverse density of partons (D0-branes) is strictly bounded to
about one per transverse Planck area. In other words the partons form
a kind of incompressible fluid. We assume that an analogous property
holds for the IKKT matrix model.

Let $V_{p+1}$ be a large enough volume of the world-volume of the p-brane.
We choose $n$ to be
\be
n \E \frac{V_{p+1}}{l_{s}^{p+1}} \,,
\label{choice}
\ee
where $l_s=\sqrt{\alpha^\prime}$ is the string length scale.
 The physical
picture of this is that the p-brane world-volume
is constituted of $n$ cells
of volume $l_s^{p+1}$. This choice of $n$ turns out to
 give correct dependence of physical quantities on the
characteristic constants.

For $A_{\mu}^{\rm cl}$ to have a dimension of length, the constants
$g_{ab}$ in eq.~\rf{bb} should be proportional to
 $V_{p+1}^{\frac{2}{p+1}}$ and should scale with
 $n$ as $n^{-\frac{2}{p+1}}$ according to the arguments of
\cite{bss} which are based on the fact that the full Hilbert space
of the dimension $n$ is represented as the tensor product
of $(p+1)/2$ Hilbert spaces of the dimension $n^{\frac{2}{p+1}}$
each. As a result, we get
\be
[A_{\mu},A_{\nu}]\E V_{p+1}^{\frac{2}{p+1}}n^{-\frac{2}{p+1}}
\label{commX}
\ee
for the commutator.

The bosonic part of the D-instanton action is
\be S\E \frac{1}{g_sl_s^4}\tr
[A_{\mu},A_{\nu}]^2 \,, \label{Sinst}
\ee
where $g_s$ is the string
coupling constant. Now, the substitution of \eq{commX} into \eq{Sinst}
gives
\be S\E \frac{1}{g_sl_s^4}
V_{p+1}^{\frac{4}{p+1}}n^{\frac{p-3}{p+1}} \,.
\label{NSinst} \ee
Substituting our choice \eq{choice} for $n$ into \eq{NSinst}, we get
\be
S\E \frac{1}{g_sl_{s}^{p+1}}V_{p+1}\E T_pV_{p+1} \,,
\ee
{\it i.e.} the
action of p-brane = tension $\times$ volume of the world-volume.

In the previous section we have computed the effective action
for the
configuration of two antiparallel D-branes. Taking into account the
scaling law \rf{commX} which yields for $c_{ab}$:
\be
c\E V_{p+1}^{\frac{2}{p+1}}n^{-\frac{2}{p+1}}\,,
\ee
we find from eq.~\rf{1/b} that
\be
W\E n c^{\frac{7-p}{2}}b^{p-7}\E V_{p+1}l_{s}^{6-2p}b^{p-7} \,.
\label{W}
\ee
This agrees with the known result from the theory of D-branes~\cite{pol}.

\newsection{Discussion}

Most of the checks, done so far, of the proposal that the IKKT matrix
model is a nonperturbative formulation
of IIB superstring deal with description
of D-branes. Our paper is also along this line.

On the other hand, the matrix model~\rf{main} should reproduce
string perturbation theory as well.
As was argued in~\cite{ikkt}, that
if large values of $n$ and
smooth matrices $A_{\mu}^{ij}$ and
$\psi_{\alpha}^{ij}$ dominate in (\ref{main}),
the commutator can be substituted
by the Poisson bracket
\be
\left[\,\cdot\,,\,\cdot\,\right] \Longrightarrow
i \{\,\cdot\,,\,\cdot\,\}
\label{cb}
\ee
and the trace can be substituted
by the integration over parameters
$\s=(\sigma_1,\sigma_2)$:
\be
\tr \ldots \Longrightarrow \int d^2 \s \sqrt{|g(\s)|} \ldots
\label{ti}
\ee
so that
the sum over $n$ and matrix integrals in \rf{main} turn into
path integrals over a positive definite function
$\sqrt{|g(\s)|}$ and over $X^{\mu}(\s)$ and
$\psi_\alpha(\s)$:
\beq
{\cal Z} = \int\, D \sqrt{|g|}\, D  X\,
 D \psi \,\e^{i{\cal S}},
\label{Smain}
\eeq
while Eqs.~\rf{action}--\rf{sym2} turn into
\be
{\cal S}=\int d^2\sigma \left(\sqrt{|g|}\,\frac{1}{g_s(\alpha')^2} \Big(
-\frac{1}{4}\{X^{\mu},X^{\nu}\}^2
+\frac{i}{2}\bar{\psi}\Gamma^{\mu}\{X_{\mu},\psi\}\Big)
+\beta \sqrt{|g|}\right),
\label{Saction}
\ee
\bea
\delta^{(1)} \psi_{\alpha} &=& -\frac{1}{2} \sqrt{|g|}
             \{ X_\mu,X_\nu \}(\Gamma^{\mu\nu}\epsilon)_\alpha ,\non
\delta^{(1)} X^{\mu} &=& i\bar{\epsilon}\Gamma^{\mu}\psi ,
\label{Ssym1}
\eea
and
\bea
\delta^{(2)}\psi_\alpha &=& \xi_\alpha ,\non
\delta^{(2)} X^{\mu} &=& 0 .
\label{Ssym2}
\eea
Here the Poisson bracket is defined by
\beq
\{X,Y\} \equiv  \frac{1}{\sqrt{|g|}}\varepsilon^{ab}\partial_a X
\partial_b Y.  \label{PB}
\eeq
The parameters $\epsilon_\alpha$ and
$\xi_\alpha$ do not depend on $\sigma_1$ and $\sigma_2$ similar to
these in Eqs.~\rf{sym1} and \rf{sym2} which are numbers rather than
matrices.

This transition from matrices to functions of $\s$
can be formalized introducing the matrix function
\be
L(\s)^{ij}= \sum_{\m} j_\m(\s) J_\m^{ij}\,,
\label{defL}
\ee
where the index ${\bf m}=(m_1,m_2)\in {\bf Z}^2$,
while $J_\m^{ij}$ form a basis for gl${}_\infty$ and
$j_\m(\s)$ form a basis in the space of functions of $\s$.
An explicit form of $j_\m(\s)$'s depends on the topology of
the $\s$-space.
Explicit formulas are available for a sphere
and a torus.%
\footnote{For a review of this subject see
\cite{Ran92} and references therein.}

The commutators of $J_\m$'s coincide with
the Poisson brackets of $j_\m$'s at least for finite $\m$'s.
This demonstrates the equivalence between
 the group of area-preserving or
symplectic diffeomorphisms (Sdiff)
and the gauge group SU$(\infty)$ for smooth configurations.

With the aid of~\rf{defL} we can relate
matrices with functions of $\s$ by
\be
A_\mu= \int d^2 \s \sqrt{|g|} X_\mu(\s) L(\s)
\label{AvsX}
\ee
and vise versa
\be
X_\mu(\s)=\tr A_\mu L(\s) \,,
\label{XvsA}
\ee
where the consequence of the completeness condition,
\be
\tr L(\s) L(\s^\prime) = \frac{1}{\sqrt{|g|}}\: \delta^{(2)} (\s-\s^\prime)\,,
\ee
has been used. The above formulas lead for smooth configurations
to Eqs.~\rf{cb} and \rf{ti}.
The word ``smooth'' here and above means precisely that
configurations can be reduced by a gauge transformation
to the form when high modes
are not essential in the expansions~\rf{AvsX} or \rf{XvsA}.

Equations~\rf{Smain} and \rf{Saction} represent
IIB superstring in the Schield formalism with fixed
$\kappa$-symmetry~\cite{ikkt}.
At fixed $\sqrt{|g|}$ the action~\rf{Saction} is invariant only
under symplectic diffeomorphisms
\be
\delta X_\mu = - \{ X_\mu, \Omega\}\,,~~~~~
\delta \psi_\alpha = - \{ \psi_\alpha, \Omega\}\,,
\ee
(in the infinitesimal form). This is an analogue of
the gauge transformation~\rf{gauge} in the matrix model~\rf{main}
at fixed $n$ which itself plays the role of $\sqrt{|g(\s)|}$.
The full reparametrization invariance of the string is
restored when one integrates over $\sqrt{|g(\s)|}$, which is an
analog of the summation over $n$ in~\rf{main}.
The matrix-model formulation is extremely nice from the point of
view of fixing the symmetry under symplectic diffeomorphisms
since this can be done by a standard procedure of fixing the gauge
in gauge theory.
It is that made it possible to calculate brane-brane
interaction by doing the one-loop calculation in the
IKKT matrix model.

A question arises whether or not these two procedures of fixing
the symmetry would always give the same result or, in other
words, are these two groups equivalent at the quantum level.
An answer to this question depends on what configurations
are essential in quantum fluctuations,
and hence what modes are essential
in the expansions~\rf{AvsX} and \rf{XvsA}. The answer to this question
is known for a pure bosonic string where
configurations which are not smooth  are certainly
important (at least in Euclidean space). 
They result in crumpled surfaces associated with
tachionic excitations. Since there is no tachion for superstring
at least perturbatively,
one might expect that only smooth configurations
are important in this case.

Another point of interest in superstring theory is calculation in
perturbation theory where higher orders
in the string coupling constant $g_s$ are associated
with non-trivial topologies of the parameter space.
The string perturbation theory should presumably arise
as a result of the loop expansion of the matrix model.
In the above language of the relations~\rf{AvsX} and \rf{XvsA}
the higher terms of string perturbation theory could
be perhaps associated with a non-trivial choice
of the basis functions $j_\m$'s corresponding to a given topology.
The algebra of symplectic diffeomorphism for non-trivial topologies
was studied and, in particular, the presence of central charges
was discovered for torus~\cite{FI88} and higher genera~\cite{BPS88}.
Analogously, it was discussed that the large $N$ limit of SU$(N)$
is not unique~\cite{PS89,HS90} and central extensions are possible.
This fact might be of interest for investigations of the matrix model.

The central point in the IKKT approach is the presence
of the $\beta$ in \rf{action} (and correspondenly in
\rf{Saction}). As is well known, Schield strings are
tensionless for $\beta=0$, and the string tension is
proportional to $\sqrt{\beta}$.

A very interesting idea of
Ref.~\cite{ikkt} is that $\beta\neq0$ can appear dynamically
in the Eguchi--Kawai reduced ten-dimensional super Yang--Mills
theory specified by%
\footnote{For a review of the reduced models see~\cite{Das87}
and references therein.}
\beq
Z_{\rm EK}= \int d\A \,d\Psi \, \e^{iS}
\label{EKmain}
\eeq
with the action
\beq
S_{\rm EK}=\frac{N a^4}{g_0^2}\left( \frac{1}{4}\tr[\A_{\mu},\A_{\nu}]^2
            +\frac{1}{2}\tr (\bar{\Psi}
           \Gamma^{\mu}[\A_{\mu},\Psi])\right)
\label{EKaction}
\eeq
where the $N\times N$ matrices $\A_\mu$ and $\Psi_\alpha$ have,
respectively, the dimension
of [mass] and [mass]${}^{3/2}$, $g_0^2$ is dimensionless and
$a$ is a cutoff. In addition to the
gauge symmetry~\rf{gauge},
the model possesses the symmetry
\be
\delta \A_\mu = \alpha_\mu \, {\bf 1}_N
\label{U1d}
\ee
whose ten parameters $\alpha_\mu$ depend on direction. 
This symmetry is crucial
for vanishing of the averages of the type
\be
\left\langle \frac 1N \tr \A_\mu \right\rangle \equiv
Z_{\rm EK}^{-1}\int d\A \,d\Psi \, \e^{iS} \frac 1N \tr \A_\mu  =0
\ee
with the integrand being invariant under the
SU$(N)$ gauge transformation~\rf{gauge} but not under~\rf{U1d}.
The symmetry~\rf{U1d} is unbroken in the perturbation
theory due to supersymmetry%
\footnote{This fact was first advocated in~\cite{MK83}
for four dimensions.}
so that there is no need of quenching or twisting in contrast
to large~$N$ QCD.

A mechanism of how the string perturbation
theory could emerge in the reduced matrix models was discussed by
Bars~\cite{Bar90}. It is based on the $1/N$ expansion of the reduced
model which leads, as for any matrix model,
to the topological expansion according to
general arguments by 't~Hooft~\cite{Hoo74}.
In order for contributions of higher genera not to be
suppressed at large~$N$, a kind of the double scaling limit
is needed,
which assumes usually fine tuning of the parameters.
It would be very interesting to find out whether or not this
mechanism works for the reduced model~\rf{EKmain} and
whether or not the double
scaling procedure suggested in~\cite{ikkt} could provide
this.

It was proposed in~\cite{ikkt}, that the term with $\beta$
in~\rf{action} which is not present in~\rf{EKaction} can
be generated in the reduced model within loop expansion.
The problem of constructing
loop expansion in the reduced model around plane vacuum,
given by the classical solution~\rf{D-instanton}, resides
in zero modes of the fermionic matrix which exist due
to the supersymmetry~\rf{sym2}.
It is still an open problem to show how the integral over
the fermionic zero modes becomes nonvanishing.

We would like to speculate on  a potential way
to deal with this problem which is related to the fact that
the integral over bosonic zero modes
\be
\int \prod_{i=1}^N d^{10} p_\mu^{(i)} =\infty
\ee
is formally divergent if $p_\mu$'s are not quenched.
Therefore, an uncertainty of the type $\infty \cdot 0$
appears in nonregularized theory which is to be done.
A useful hint on how the result can look like is given by
the simplified model~\cite{KSS96} where the
partition function was calculated via the Nicolai map.

There exists one more potential way out for the problem
of the fermionic zero modes --- the same as in superstring
theory ---
where simplest amplitudes are well-known to
vanish exactly for the same reason.
One should consider in the reduced model averages of
several operators (analogous to vertex operators in superstring
theory) to make the integral over the fermionic zero modes
nonvanishing.

\subsection*{Acknowledgments}

We thank A.~Gorsky, A.~Mikhailov,
A.~Morozov and P.~Olesen for useful discussions.
This work was supported in part by INTAS grant 94--0840 and by
CRDF grant 96--RP1--253. Y.M. was sponsored in part by the Danish Natural
Science Research Council.
The work of K.Z.\ was supported in part by RFFI grant 96--01-00344.

\subsection*{Note added}

After this paper was submitted for publication,
the revised version of~\cite{Tse97} has appeared
which proposes to interpret the classical solutions
in the IKKT matrix model as bound states
of D-branes with large (of order $n$) number of
D-instantons, in analogy with previous work~\cite{LM96}
on the BFSS matrix model.

\eop

\end{document}